# USING A COCONTRACTION RATIO TO PREDICT ANTAGONISTIC BEHAVIOR DURING ELBOW MOTION

[1,2] Charles Pontonnier and [1,3] Georges Dumont
[1] Bunraku project team, IRISA, Campus de Beaulieu, 35042 Rennes Cédex, France. charles.pontonnier@irisa.fr
[2] Department of Health Science and Technology, Frederik Bajers Vej, DK-9220 Aalborg, Denmark
[3] ENS Cachan Antenne de Bretagne, Campus de Ker Lann, F-35170 Bruz, France

## SUMMARY
Inverse dynamics methods for muscle forces prediction are globally unable to predict antagonistic activity during a joint motion. This is due to a lack of physiological information describing how forces are shared between flexors and extensors. The aim of this study is the definition and the use of a new EMG-based cocontraction ratio in an inverse dynamics muscle forces prediction approach applied to the elbow flexion motion. Results show the relevance of the ratio.

## INTRODUCTION
The mechanical impedance of each joint of the human body is controlled by the cocontraction of both agonistic and antagonistic muscles.
Cocontraction can be generated when muscles are involved in multiple joint motions [1]. Mathematically, cocontraction in a single joint cannot be predicted with standard inverse dynamics approaches [2]. It also seems necessary to add physiological information in the optimization problem used to predict muscle forces.
Using experimental data has been proposed in order to predict cocontraction. EMG measurement can be used to frame the activation of antagonistic muscles [3]. This solution leads to realistic contraction patterns, but it is subject to scaling problems and muscle model or experimental uncertainty (such as Maximum Voluntary Contraction MVC). We propose to use a ratio instead of a direct activation value to obtain antagonistic activation and to avoid uncertainties on MVC value. Our hypothesis is to consider that dynamics of the motions lead to specific cocontraction patterns that can be used to predict the muscle forces. These cocontraction patterns can only be obtained by using EMG measurement during the motion. In the elbow joint case, triceps is globally antagonistic during flexion, and also during extension for a neutral shoulder position. We will use this muscle in application of this method.

## METHODS
Using a framework and a musculoskeletal model of the arm fully described in [4, 5], we process motion capture and EMG data to obtain muscle forces (Figure 1). We capture the trunk and right arm motion. For the elbow joint we use bipolar EMG measurement of Biceps and Triceps activity to build the cocontraction ratio, assuming that the behavior of flexors is close to the biceps one and behavior of extensors is close to the triceps one. EMG signals are processed in the same way as proposed in [6] to obtain muscle activations.

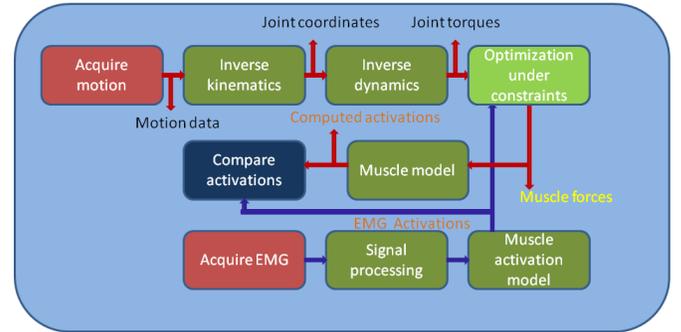

**Figure 1**: Inverse dynamics framework using both motion capture and EMG data to compute muscle forces.

Five muscles are taken into account in our elbow joint model: Biceps, Triceps, Brachialis, Brachioradialis, Anconeus. Standard Hill-type models [7] are associated with these muscles and parameters are adjusted according to [8].

The optimization problem is defined at each frame as follows :

$$\min f = \sum_{k=1}^{n}\left(\frac{F_k(i)}{F_k^{\max}(i)}\right)^2$$

under constraints :

$$(a) \quad \Gamma(i) = \sum_{k=1}^{n} F_k(i) R_k(i) \qquad (3)$$

$$(b) \quad F_k^{\min}(i) \leq F_k(i) \leq F_k^{\max}(i)$$

$$(c) \quad k_{global}(i)\cdot(1-\gamma) \leq \frac{F_{biceps}(i)}{F_{triceps}(i)} \leq k_{global}(i)\cdot(1+\gamma)$$

The constraints are (a) mechanical equilibrium of the joint (b) Physiological limits of the muscles (c) cocontraction constraint (with a confidence indice γ of 0.1).
The cocontraction ratio for the elbow flexion/extension joint $k_{global}(i)$ is defined at each frame i as follows:

$$k_{global}(i) = \frac{F_{biceps}^{EMG}(i)}{F_{triceps}^{EMG}(i)} \qquad (2)$$

Where $F_{biceps}^{EMG}(i)$ and $F_{triceps}^{EMG}(i)$ are computed from the muscle models and the EMG-based activations measured for the frame i.

Experimental data has been collected on 4 healthy right handed male subjects without any specific training. They performed each 40 elbow flexion/extension motions at 10 different speeds (Figure 2) and 4 different loads (barbells). In order to minimize the muscle fatigue effects, each subject realizes motions a randomized order, and 5 minutes pauses are managed every 5 motions, allowing the subject to rest, drink and eat if he needs to. Most restricting motions can be followed by a pause, on subjects request.

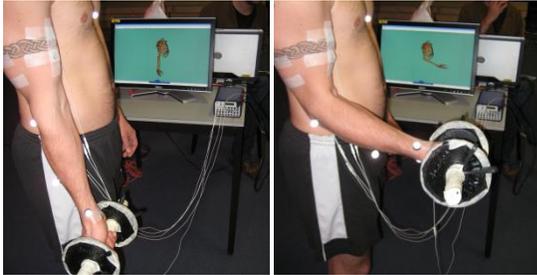

**Figure 2**: Experimentation situation. The subject follows a visual guide to perform motions.

## RESULTS AND DISCUSSION

The influence of the cocontraction constraint on computed forces is studied by comparing the EMG activations with computed activations, using a global scale factor to decrease calibration issues (Isometric maximum forces, MVC).

Figures 3 and 4 show the mean error for the whole subjects between measured and computed activation without and with co-contraction constraint for the 40 motions. Each bar represents the error made on a motion (defined by a speed and a load) for the whole 4 subject.

Without any co-contraction information, the estimation is absolutely unrealistic. The addition of the co-contraction constraint leads to a decrease of 70% the error made on the estimation. Even if the cocontraction ratio seems to be relevant to predict the behavior of antagonistic muscles, uncertainties on muscle parameters and calibration issues leads to an important remaining error.

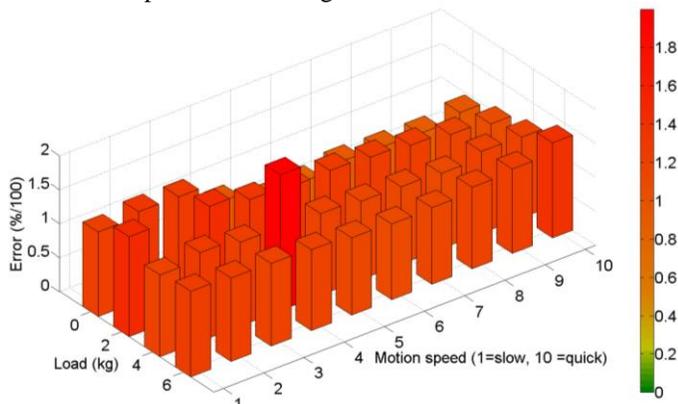

**Figure 3**: Error made on triceps activation without cocontraction constraint.

## CONCLUSION

The addition of a cocontraction constraint in the optimization problem of an inverse dynamics approach leads to a more realistic behavior of antagonistic activations. Global variations of the activity are well predicted, but it still remains a static error due to the numerous arbitrary parameters introduced in muscle models. In order to improve these first results, calibration of muscle models and MVC procedures has to be more precise.

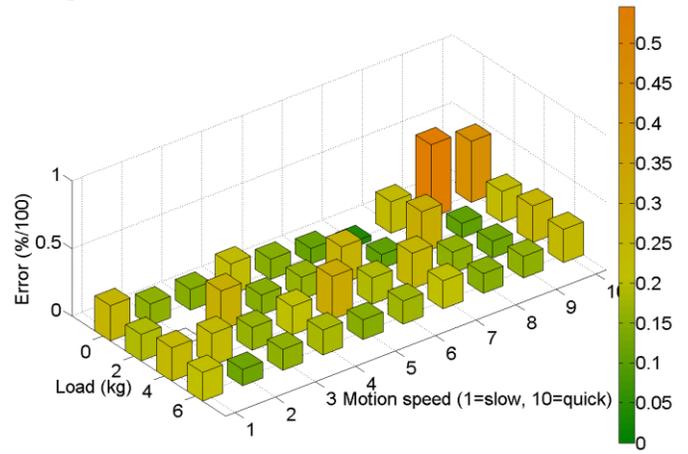

**Figure 4**: Error made on triceps activation with cocontraction constraint.

## REFERENCES

1. Herzog W., and Binding P., Cocontraction of pairs of antagonistic muscles : analytical solution for planar static nonlinear optimization approaches, *Mathematical Biosciences*, **118(1)** : 83 – 95, 1993.
2. Ait-Haddou R. et al., Theoretical considerations on cocontraction of sets of agonistic and antagonistic muscles, *Journal of Biomechanics*, **33(9)** :1105 – 1111, 2000.
3. Vigouroux L. et al., Estimation of finger muscle tendon tensions and pulley forces during specific sports-climbing grisp techniques, *Journal of Biomechanics*, **39(14)** :2583 – 2592, 2006.
4. Pontonnier C. and Dumont G., Inverse dynamics method using optimization techniques for the estimation of muscle forces involved in the elbow motion, *IJIDeM,* **3**: 227 – 236, 2009.
5. Pontonnier C. and Dumont G, From motion capture to muscle forces in the human elbow aimed at improving the ergonomics of workstations*, Virtual and Physical Prototyping*, **5(3)**: 113 – 122, 2010.
6. Buchanan S. et al., Neuromusculoskeletal modeling : estimation of muscle forces and joints moments and movements from measurement of neural command, *Journal of Applied Biomechanics*, **20** : 367-395, 2004.
7. Rengifo C. et al., Distribution of forces between synergistic and antagonistic muscles using an optimization criterion depending on muscle contraction behavior, *Journal of Biomechanical Engineering*, **132**, 2010.
8. Pontonnier C. and Dumont G., Motion analysis of the arm based on functional anatomy, Proceedings of 2$^{nd}$ Workshop on 3D Physiological human, *LNCS*, **5903**, Zermatt, Switzerland, 2009.